\documentclass{article}

\usepackage{xspace}
\usepackage[plain]{algorithm}
\usepackage{framed}
\usepackage{float}
\usepackage{placeins}
\usepackage{cite}
\usepackage[noend]{algpseudocode}
\usepackage{graphicx}
\usepackage{subcaption}
\usepackage{tikz}
\usetikzlibrary{arrows.meta,calc,positioning,shapes.geometric}
\usepackage{authblk}

\usepackage{amsmath}
\usepackage{amsthm}
\usepackage{amssymb}
\usepackage{url}

\usepackage{xcolor}

\usepackage{comment}

\algblockdefx{AtState}{END}[1]{\underline{In state {\sf #1}:}}{}
\algblockdefx{AtStateBreakLine}{END}[2]{ \underline{In state {\sf #1}} \newline  \underline{{\sf #2}:} }{}
\algcblockdefx{AtState}{AtState}{END}[1]{\underline{In state {\sf #1}:}}{\vspace{-1.5em}}{}
\algcblockdefx{AtStateBreakLine}{AtStateBreakLine}{END}[2]{ \underline{In state {\sf #1}} \newline  \underline{{\sf #2}:} }{}

\newcommand{\G}{\mathcal{G}}

\newcommand{\thickhline}{%
    \noalign {\ifnum 0=`}\fi \hrule height 1pt
    \futurelet \reserved@a \@xhline
}

\newcommand{\BH}{{\sc Bh}\xspace}
\newcommand{\BHS}{{\sc Bhs}\xspace}

\newtheorem{lemma}{Lemma}
\newtheorem{theorem}{Theorem}
\newtheorem{definition}{Definition}
\newtheorem{observation}{Observation}

\newcommand{\ic}{\ic}

\newcommand{\tc}{\tc}
\newcommand{\R}{\textnormal{\textsc{Retroguard}}\xspace}

\newcommand{\A}{\textnormal{\textsc{Avanguard}}\xspace}
\newcommand{\Leader}{\textnormal{\textsc{Leader}}\xspace}
\newcommand{\JointLeader}{\textnormal{\textsc{Joint Leader}}\xspace}
\newcommand{\AggressiveLeader}{\textnormal{\textsc{Aggressive Leader}}\xspace}
\newcommand{\Scout}{\textnormal{\textsc{Scout}}\xspace}
\newcommand{\Forward}{\textsf{Forward}\xspace}
\newcommand{\CautiousWalk}{\textsf{CautiousWalk}\xspace}
\newcommand{\RT}{\textsf{CautiousPendulum}\xspace}
\newcommand{\BackwardCP}{\textsf{BackwardCP}\xspace}

\newcommand{\GL}{\textsf{Gather\&Locate}\xspace}

\newcommand{\JointCW}{\textsf{JointCW}\xspace}

\tikzset{
  glnode/.style={circle,draw,line width=.55pt,fill=white,
    minimum size=4.2mm,inner sep=0pt},
  glagent/.style={rectangle,draw,line width=.55pt,fill=white,
    minimum width=4.8mm,minimum height=4.8mm,
    inner sep=.2pt,font=\scriptsize\sffamily},
  glgroup/.style={draw,rounded corners=1.5pt,line width=.55pt,
    fill=gray!5,align=center,inner sep=2.5pt,font=\scriptsize},
  glpebble/.style={circle,fill=black,minimum size=1.6mm,inner sep=0pt},
  glmove/.style={-{Latex[length=1.6mm,width=1.1mm]},line width=.6pt},
  glrmove/.style={glmove,densely dashed},
  glbimove/.style={{Latex[length=1.6mm,width=1.1mm]}-{Latex[length=1.6mm,width=1.1mm]},
    densely dashed,line width=.6pt},
  glbox/.style={draw,rounded corners=1.5pt,line width=.55pt,
    fill=gray!5,align=center,inner xsep=3pt,inner ysep=2.5pt,
    font=\scriptsize},
  gledge/.style={line width=.55pt}
}

\title{Black Hole Search by Scattered Agents in Dynamic Rings} 

\author[1]{Giuseppe A. Di Luna}
\author[2]{Paola Flocchini}
\author[3]{Giuseppe Prencipe} 
\author[4]{Nicola Santoro}
\affil[1]{DIAG, Sapienza University of Rome, Italy}
\affil[2]{School of Electrical Engineering and Computer Science, University of Ottawa, Canada}
\affil[3]{Dipartimento di Informatica, Universit\`a di Pisa, Italy}
\affil[4]{School of Computer Science, Carleton University, Canada}

\begin{document}

\maketitle

\begin{abstract}
We study \emph{Black Hole Search} (\textsc{Bhs}) in synchronous dynamic rings
by mobile agents that start \emph{scattered}, i.e., at distinct arbitrary
nodes. A black hole is a stationary node that silently destroys every agent
that enters it, leaving no detectable trace. An algorithm solves \textsc{Bhs}
if at least one agent survives and terminates knowing the footprint of the
ring, that is, its own position relative to the black hole; every agent that
terminates must have this knowledge.

The ring is oriented and \emph{$1$-interval connected}: in each round an
adversary may remove one edge, subject only to the graph remaining connected.
 We give a complete characterization
of size-optimal solutions. As in the co-located case, two agents cannot solve
\textsc{Bhs}; we therefore focus on three agents. We present
\textsc{Gather\&Locate}, an algorithm in the \emph{pebble} model in
which three scattered agents solve \textsc{Bhs} using $O(n^2)$ moves and
$O(n^2)$ rounds. This is asymptotically optimal: we prove a matching
$\Omega(n^2)$ lower bound on both moves and rounds that holds for every
three-agent algorithm even in the stronger \emph{whiteboard} model. Since two static-ring agents locate a black
hole in $O(n)$ moves, and three co-located agents suffice in $O(n^{1.5})$ on
dynamic rings, our results quantify the combined cost of dynamicity and
scattering.

Finally, we show that an exogenous marking mechanism is necessary: in the
\textsc{FaceToFace} model, where agents communicate only upon
meeting, three scattered agents cannot solve \textsc{Bhs} even on a static
oriented ring, in contrast with the co-located case.
\end{abstract}


\section{Introduction}

\subsection{Exploration of Dynamic Networks}

In the field of distributed computing, a substantial body of research (see \cite{FlPS19}) has focused on the computational paradigm of {\em mobile agents}. A {\em mobile agent} is a software entity capable of navigating a network by visiting nodes. Upon visiting a node, the agent executes computations while interacting with the local environment, including the memory and resources of the node, before transmitting itself to a neighboring node. Essentially, a mobile agent can be regarded as an intelligent message with computational abilities, capable of determining its next destination.

Numerous problems have been studied under the mobile agent paradigm. Notable examples include: {\em exploration}, where a team of agents must collectively traverse the entire network; {\em gathering}, where agents converge at a common, initially unknown node; and {\em patrolling}, where agents periodically monitor the network, minimizing the interval between consecutive visits to the same node.

One well-studied problem in this domain is the Black Hole Search (\BHS) \cite{MaS19}. This involves identifying a dangerous stationary node, termed the {\em black hole} (\BH), which silently destroys any agent that visits it. A \BH models various network failures, such as a crashed host that eliminates visiting agents or a virus-infected node that deletes incoming messages. Extensive research on the \BH problem has provided a comprehensive understanding of its computational properties under different assumptions, including communication mechanisms, synchronicity levels, network topology, and agent capabilities. However, most studies have considered networks to be {\em static}, where nodes and links remain unchanged.

Recently, distributed computing research has shifted towards studying mobile agents in {\em highly dynamic graphs}, where topological changes occur frequently and are not limited to sporadic failures or congestion. Such graphs represent various modern networks whose dynamic nature arises from advancements in communication technologies (e.g., wireless networks), software architectures (e.g., software-defined network controllers), and societal changes (e.g., ubiquitous smart mobile devices) (see \cite{CaFQS12}).

We investigate {\em evolving graphs}, a class of dynamic graphs represented as infinite sequences of static graphs. In this model, computation is inherently {\em synchronous}, with each round corresponding to a static graph in the sequence. A widely adopted assumption in this setting is {\em 1-interval connectivity}, which ensures that the graph remains connected at every round (e.g., \cite{AbsM14,briefdiluna,HaeK12,KuhLyO10,KuhLoO11}).

Our focus is on 1-interval connected rings, where the network topology is a ring graph where, in each round, an edge might be missing.

In recent years, the study of agents on highly dynamic graphs has grown rapidly (see the survey in \cite{DiL19}). Significant progress has been made on 1-interval connected rings, with investigations into the {\em gathering} problem \cite{DiLFPPSV20}, the {\em exploration} problem \cite{DiLDFS20, BournatDD16,BournatDP17}, and the \BHS for co-located\footnote{Agents are co-located when they all start from the same node.} agents \cite{DiFPS25}. 

Despite this progress, many questions remain open. This paper addresses one such question: how does the computational complexity of finding a \BH change when agents are {\em scattered} ? In the scattered scenario, agents start from arbitrary nodes in the graph. We demonstrate that this setting differs significantly from the co-located case (studied in \cite{DiFPS25}) in terms of both solvability (some scenarios become unsolvable) and computational complexity.

\subsection{Related Works}

The Black Hole Search (\BHS) problem  
has ben introduced in \cite{Dobrev2002}. The problem has been studied
in graph of restricted topologies (e.g., trees \cite{CzKMP07}, rings and tori \cite{dobrev2007locating,opodis12,ChDLM13})
and  in  arbitrary and possibly unknown topology (e.g., \cite{Dobrev2002,dobrev2013exploring,CzKMP06}). 
For a recent survey see  \cite{MaS19}.
 The most relevant papers are the ones investigating  \BHS\ in static ring networks. 
In the {\em asynchronous} setting, it is possible to solve
    the problem with two  co-located agents and   $\Theta(n \log n)$   moves,
   in the Whiteboard model~\cite{DobrevFPS07},
and in the Pebble model~\cite{FIS12}.
It has been shown that  ${\cal O}(n \log n)$ moves also suffice for the scattered case and oriented rings \cite{ChDLM13}. 
Others~\cite{BaFMS11} investigated time-optimal algorithms when every
edge traversal takes one time unit.

%

 In spite of all the differences in settings and assumptions, 
all these investigations share a common trait: 
the agents operate on a {\em static} network.

The only works studying \BHS in dynamic graph are \cite{DiFPS25} and \cite{FlKMS12}.
   \cite{FlKMS12} is on the black hole search in {\em carrier graphs}, a particular class of periodic temporal graphs 
defined by circular intersecting routes  of public carriers, where the stops are the nodes of the graph and  
the agents can board and disembark from a carrier at any stop.

In~\cite{DiFPS25}, the authors study the same dynamic-ring model but
with initially co-located agents. They show that three agents are
necessary to solve \BHS\ and give optimal algorithms using three agents
(in a static graph, two agents suffice to explore any known graph
\cite{CzKMP06}). With communication only between co-located agents,
their algorithm uses \(\Theta(n^2)\) moves and rounds; with pebbles, it
uses \(\Theta(n^{1.5})\) moves and rounds.
  
 In both studies, the agents are assumed to be initially {\em co-located}, i.e.,
 to start from the same node, which is guaranteed not to be the \BH.
To the best of our knowledge, no study considers the case of scattered agents.

\subsection{Contributions}

We study  the problem of
finding a \BH\ in an oriented dynamic ring by scattered agents with
distinct visible identifiers under different communication capabilities.

We study the exogenous communication models defined in
Section~\ref{sec:model}: the Pebble model and the Whiteboard model.

We focus on \emph{size-optimal} algorithms, namely algorithms that use
the minimum number of agents needed to solve the problem in the given
communication model. It is known that two agents cannot solve \BHS\ in
dynamic rings~\cite{DiFPS25}; the same impossibility holds in the
scattered case.
Therefore, we consider size-optimal algorithms using three agents.
In Theorem~\ref{scat:lb}, we show that any size-optimal algorithm solving
\BHS\ requires \(\Omega(n^{2})\) moves and \(\Omega(n^{2})\) rounds in
the Whiteboard model.
On a static synchronous ring, two agents can find the black hole in
\(O(n)\) moves and rounds. Thus, dynamicity not only increases the number
of agents required but also significantly increases the move and round
complexity. This comparison also highlights the cost of scattering the
agents: on dynamic rings, \(O(n^{1.5})\) moves and rounds suffice when the
agents are co-located~\cite{DiFPS25}.
Finally, our lower bound is tight: we provide an algorithm that solves
\BHS\ in the Pebble model using \(O(n^2)\) rounds and moves with three
agents (Theorem~\ref{th:scatfinal}).\footnote{{\color{red}} dire qui che pebble è più debole ed è quindi confrontabile con whiteboard}

We conclude by showing that a communication mechanism that can mark nodes,
such as pebbles or whiteboards, is necessary for a three-agent solution. In
fact, in the FaceToFace (F2F) model, where agents can communicate only when
they occupy the same node, three agents cannot solve \BHS\ even on a static
ring of size \(30\) (Theorem~\ref{th:three-f2f-static}).

%

\section{Model and Preliminaries}

\subsection{The Model and the Problem}\label{sec:model}

We number rounds from \(1\). The system is a synchronous evolving graph
\(\G=(G_1,G_2,\
)\) on a fixed node set \(V\). Time is divided into
rounds, and \(G_r=(V,E_r)\) contains exactly the edges present in round
\(r\). The \emph{footprint} of \(\G\) is the static graph
\((V,\bigcup_{r\geq 1}E_r)\). The evolving graph is \emph{1-interval
connected} if \(G_r\) is connected in every round \(r\).

We study \emph{dynamic rings}: 1-interval connected evolving graphs
whose footprint is a ring. Let
\({\cal R}=(v_0,v_1,\ldots,v_{n-1})\) be an oriented dynamic ring,
where indices are modulo \(n\). The right, or clockwise, port of
\(v_i\) leads to \(v_{i+1}\), and the left, or counter-clockwise, port
leads to \(v_{i-1}\). Nodes are anonymous and therefore have no
identifiers.

The ring contains one \emph{black hole} (\BH) node, which destroys every
agent that enters it without leaving detectable evidence. Every other
node is called \emph{safe}. A set
\(A=\{a_0,a_1,\ldots,a_{k-1}\}\) of mobile agents starts at distinct,
arbitrarily chosen safe nodes; such an initial placement is called
\emph{scattered}. Agents occupying the same node are \emph{co-located}.
The agents have distinct visible identifiers in
\(\{0,\ldots,k-1\}\), know \(n\), and have bounded internal memory ,
\({\cal O}(\log n)\) bits suffice for our algorithms.

All agents are activated in every round. Upon activation at a node
\(v\) in round \(r\), an agent takes a local snapshot containing the
agents and pebbles ({\bf GP: if any}) at \(v\), together with the presence or absence of
its two incident edges.
It then performs the communication step allowed by the communication
model, updates its local memory, and chooses to terminate, to stay, or
to attempt the traversal of one incident edge.

A terminated agent performs no further actions. If, in round \(r\), an
active agent successfully attempts to traverse \((v,v')\in E_r\), it
reaches \(v'\) at the beginning of round \(r+1\); this successful
traversal is one \emph{move}. If that edge is absent, the agent remains
at \(v\) and makes no move. An agent is
\emph{surviving} if it has not been destroyed by the \BH.

An \emph{execution} is determined by an initial placement of the
agents, the position of the \BH, a sequence of edge sets
\((E_r)_{r\geq 1}\), and the actions prescribed by an algorithm. All
algorithms considered in this paper are deterministic.

We consider two \emph{exogenous communication models}, in which agents
exchange information through pebbles or whiteboards:

\noindent - \emph{Pebble:} each agent initially carries one movable
pebble. An agent may place any pebble it carries and may take any pebble
at its current node, including one placed by another agent; thus an
agent may carry more than one pebble. Concurrent actions that place or
take pebbles at the same node are performed in fair mutual exclusion.

\noindent - \emph{Whiteboard:} each node contains a shared memory of
size. Agents at the node may read and write this memory.
All whiteboards initially contain the same blank value.
Access is governed by \emph{fair mutual exclusion}\footnote{Multiple
agents may read the same whiteboard in one round. If several agents try
to write simultaneously, only one succeeds; fairness means that every
agent that continues trying eventually succeeds.}.

In both models, co-located agents see one another's visible identifiers
and may exchange constant-size messages during the communication step.
The messages used by our algorithms communicate, for example, the
sender's current role.

We remark that the  stronger whiteboard model  will  be used to prove lower bounds.
On the other hand, all our algorithms are designed for the weaker model of Pebbles.

An agent \emph{knows the footprint} of \({\cal R}\) if it knows the
number of ring edges from its current node to the \BH\ in both the
clockwise and counter-clockwise directions. This information lets the
agent reconstruct the ring and its position relative to the \BH.

\begin{definition}(\BHS) \cite{DiFPS25}
An algorithm \({\cal A}\) solves \BHS\ on a dynamic ring \({\cal R}\)
if at least one agent survives and terminates knowing the footprint of
\({\cal R}\), and every agent that terminates knows that footprint.
\end{definition}

Since \(n\) is known, either one of these two distances determines the
other; the agent need not visit every node of \({\cal R}\).

The \emph{size} of an algorithm is the number of agents it uses. For an
execution in which an agent terminates knowing the footprint, its
\emph{time} is the number of rounds through the first such termination,
and its \emph{cost} is the total number of moves made by all agents
through that termination. The time and cost of an algorithm are the
worst of these values over all permitted executions. An algorithm is
\emph{size-optimal} in a
communication model if no algorithm in that model solves \BHS\ with
fewer agents.

\section{Preliminaries}\label{sec:preliminaries}

Before presenting and analyzing our algorithms, we report some known
impossibility results and a technical observation that will be used later.
Also, we briefly describe a well known 
idea employed for \BHS\ in static graphs that will be adapted in our algorithms, as well as
the conventions and symbols that will be used in our algorithms.

\subsection{Known Impossibilities and Basic Definitions}

In this section we report known impossibility results and basic definitions. 

\begin{theorem}\cite{DiFPS25}\label{madebyus0}
In a dynamic ring    of size $n>3$,
two co-located agents cannot solve \BHS.
%
The impossibility holds even if the agents have distinct visible identifiers
and use the Whiteboard model.
\end{theorem}

 \begin{theorem}\cite{DiFPS25}
There exists no algorithm that solves the \BHS\ in a dynamic ring  ${\cal R}$  whose size is unknown to the agents.
The result holds even if the nodes have whiteboards, the agents have
distinct visible identifiers, and arbitrarily many agents are available.
\end{theorem}

\begin{definition} (Blocked Agent) An agent is blocked in round \(r\)
if it attempts to traverse an incident edge that is absent in that
round.
\end{definition}
 
\begin{observation}\cite{DiFPS25}\label{madebyus}
Let \(U \subsetneq V\) be a connected set consisting of at least two
consecutive nodes in the footprint of a dynamic ring \({\cal R}\). Let
\(e_c\) and \(e_{cc}\) denote, respectively, the clockwise and
counterclockwise boundary edges between \(U\) and \(V \setminus U\).
Suppose that all agents are located in \(U\) at the beginning of round
\(r\). An algorithm can guarantee that at least one agent reaches
\(V \setminus U\) in round \(r+1\) only if, in round \(r\), one agent
can traverse \(e_c\) and another agent can traverse \(e_{cc}\).
\end{observation}

\begin{proof}
Suppose that the stated condition does not hold. Then, in round \(r\),
at most one of the two boundary A valid dynamic-ring schedule can omit it in round
\(r\). Consequently, no agent can leave \(U\), and hence the algorithm
cannot guarantee that an agent reaches \(V \setminus U\) in round
\(r+1\).
\end{proof}

\subsection{The \CautiousWalk\ primitive}\label{sec:cautious-walk}

The \CautiousWalk\ primitive, introduced in~\cite{Dobrev2002}, lets an agent
test the safety of a neighbouring node while preventing a second agent
from entering the \BH\ through the same edge. Suppose that an agent
\(a\) is at a node \(u\) known to be safe and must test its neighbour
\(v\). While \((u,v)\) is absent, \(a\) waits and keeps carrying its
pebble. In the first round in which the edge is present, \(a\) leaves
its pebble on \(u\) and crosses \((u,v)\). Until \(a\) returns, no other
agent at \(u\) crosses that edge. If \(v\) is safe, \(a\) returns to
\(u\), takes its pebble,
and crosses \((u,v)\) again. We call these three movements --- the first
crossing, the return, the final crossing --- one \emph{cautious
step}. After the step, \(v\) is known to be safe and becomes \(a\)'s
new current node.

\subsection{\RT: A procedure for co-located agents}
\label{sec:cautious-pendulum}

We also use as a primitive the procedure \RT\ of~\cite{DiFPS25}. It identifies the \BH\ in a
1-interval connected dynamic ring using three co-located agents.
It starts with three co-located agents having distinct visible
identifiers and assigns them the roles \Leader, \A, and \R. Essentially, \Leader and \A move clockwise simulating a cautious walk, while \R moves counterclockwise to explore new nodes in that direction.

At a safe node \(u\), \Leader\ and \A\ test the clockwise neighbour
\(u^+\). If \((u,u^+)\) is absent, \A\ waits. When the edge is present,
\A\ crosses to \(u^+\). If \A\ survives, it returns to \(u\) the next
time the edge is present; \Leader\ and \A\ then cross to the now known
safe node \(u^+\). More precisely, once \A\ is at \(u^+\), if
\((u,u^+)\) is present in round \(r\) but \A\ does not reach \(u\) at
the beginning of round \(r+1\), \Leader\ terminates knowing that
\(u^+\) is the \BH.

Whenever \R\ leaves the node occupied by \Leader\ and moves 
counter-clockwise, we call that activation a \emph{departure}. On its
\(i\)-th departure, \R\ moves counter-clockwise past the \(i-1\) nodes
tested on its earlier departures and attempts to enter the next node.
This attempt is the \emph{test} of that node. If \R\ survives, it
reverses direction and moves clockwise until it is again co-located
with \Leader. We call that co-location after a departure a
\emph{report}. After a report, \R\ starts its next departure. If an
edge on this route is absent, \R\ waits for it to reappear.
 

To distinguish a delayed report from a missing report, after each
departure \Leader\ counts its blocked clockwise attempts. Whenever such
an attempt is blocked, \Leader's clockwise edge is absent and every
other ring edge is present, so a surviving \R\ makes progress along its
route. If \R\ has not reported after \(2n\) blocked attempts by
\Leader, it must have entered the \BH\ at the node tested on that
departure; \Leader\ can therefore terminate knowing the location of
the \BH. We call this rule the \emph{failed-report test}.
The procedure \BackwardCP, defined in
Section~\ref{scatter:threeagents}, uses the same test.

\begin{theorem}\label{th:pendulum} \cite{DiFPS25}
  Consider a dynamic ring \({\cal R}\) with three co-located agents with distinct visible identifiers. 
  Procedure \RT solves  \BHS\  with ${\cal O}(n^2)$ moves and in ${\cal O}(n^2)$ rounds. 
\end{theorem}

\section{Lower Bound for Size-optimal Algorithms}
We now prove a quadratic lower bound on the moves and rounds of every
size-optimal \BHS\ algorithm for scattered agents, in the Whiteboard model.
The bound holds even
with distinct visible identifiers and whiteboards.

\begin{theorem}\label{scat:lb}
In the Whiteboard model, every \BHS\ algorithm for three scattered
agents with distinct visible identifiers requires \(\Omega(n^2)\)
moves and \(\Omega(n^2)\) rounds on some dynamic ring.
\end{theorem}
\begin{proof}
 Let ${\cal A}$ be an algorithm that solves \BHS, and let $a$, $b$, and $c$ be the three agents. Suppose the agents have unique identifiers and, without loss of generality, let $c$ be the first agent to move.

Assume an initial configuration where $c$ is placed on node $v_c$, a neighbor of the \BH, and agents $a$ and $b$ are on two neighboring nodes, $v_a$ and $v_b$. The distance of the closest between $a$ and $b$ to the \BH is a value unknown to the agents, between $\frac{n}{3}$ and $\frac{(n-1)}{2}$.
Without loss of generality, place $c$ so that its first move is clockwise and it enters the \BH; thus, $v_c$ is the counter-clockwise neighbor of the \BH.

Let $U_r$ denote the set of nodes explored by agents $a$ and $b$ at the end of round $r$. By Observation~\ref{madebyus}, agents $a$ and $b$ are sure to explore a node outside $U_r$ only if they attempt to traverse, in the same round, both edges crossing the cut between $U_r$ and $V \setminus U_r$. If they do not do so, the edge that one of them tries to traverse may be blocked, and thus $U_r = U_{r+1}$.

Whenever $a$ and $b$ attempt to leave $U_r$ using both outgoing edges, choose one of these two edges as the missing edge, always choosing it so that no agent can traverse the edge leading to $v_c$. This implies that neither $a$ nor $b$ can read the message left by $c$ on $v_c$, and thus cannot discover the location of the \BH using this information.

Note also that from the above it follows that whenever $U_r \neq U_{r+1}$, agents $a$ and $b$ are at opposite ends of $U_{r+1}$.

Eventually, another agent will die by entering the \BH from the counter-clockwise direction; without loss of generality, let this be agent $b$. When this happens, the distance between $a$ and the \BH is at least $\frac{n}{3}$. From that round onward, choose the missing edge in each round so that $a$ remains on two adjacent nodes. This implies that $a$ cannot visit the nodes closer to the \BH where $b$ has left information on the whiteboards.

Thus, starting from the first round $r$ with $|U_r| = \frac{n}{3}$, whenever $U_{r} \neq U_{r+1}$, the two remaining agents are on opposite sides of $U_{r+1}$. Before the next round $s>r$ for which $U_s \neq U_{s+1}$, they must collectively traverse at least $\frac{n}{3} - 1$ nodes. Otherwise, if one of them enters the \BH in round $s$, the surviving agent neither knows the location of the \BH nor can reach a whiteboard containing this information.

If this traversal does not occur between two consecutive rounds in which the set of explored nodes grows, the remaining agent can be confined forever to two adjacent nodes and may never discover the exact location of the \BH. Since the \BH can be located at a distance between $\frac{n}{3}$ and $\frac{(n-1)}{2}$, there must be $\Omega(n)$ rounds $r$ satisfying $U_r \neq U_{r+1}$, and each pair of consecutive such rounds requires $\Omega(n)$ moves by the agents. Therefore, the total number of moves required is $\Omega(n^2)$. The bound on the number of rounds follows directly from the number of moves, noting that there are three agents.
\end{proof}

The algorithm \GL\ described in Section~\ref{scatter:threeagents}
matches this lower bound.

\section{An Optimal Algorithm: \GL}\label{scatter:threeagents}

We now present \GL, a three-phase algorithm for three scattered agents
equipped with pebbles, that matches the lower bound of the previous section.

\subsection{Algorithm description}

The algorithm is divided in three phases. Phase~1, comprising rounds
\(1,\ldots,6n\), uses clockwise cautious exploration to form one logical
group in one of two boundary configurations. Phase~2, comprising rounds
\(6n+1,\ldots,6n+T\), where \(T=4n^2\), dispatches each configuration to the appropriate
search procedure and allows any pending cautious return to finish.
Phase~3 begins in round \(6n+T+1\) and completes any execution that is
still unresolved.

At every activation, completed returns and pebble recovery are processed
before termination tests, timeouts, role changes, and movement. A group committed to an
unfinished cautious step is neither split nor reassigned before that
step is resolved.

\subsubsection{Phase 1: forming one cautious group}

Phase~1 consists exactly of rounds \(1,\ldots,6n\) and forms one logical
cautious group from the scattered agents. Each agent begins as a singleton
and repeatedly applies the clockwise \CautiousWalk\ primitive of
Section~\ref{sec:cautious-walk}. We call the pebble placed for an unfinished
cautious step its \emph{step pebble}. Agents that meet merge into a {\em group} and
execute the procedure \JointCW\ described below. If no agent terminates during
Phase~1, then at the beginning of round \(6n+1\) the surviving agents occupy
one of the configurations characterized below.

\paragraph{Procedure \JointCW.}
\noindent\emph{Roles.}
A group executing \JointCW\ contains two or three agents. Its largest-identifier
member assumes the role \A; every other member assumes the role
\JointLeader. Hence a
two-agent group has one \JointLeader, whereas a three-agent group has
two.

\smallskip
\noindent\emph{Purpose.}
The agents start scattered, so they cannot initially run the
three-agent procedure \RT. \JointCW\ lets agents that meet merge into a
single cautious group while preserving the fundamental safety rule:
only one agent probes a node whose safety is unknown.

\smallskip
\noindent\emph{Operation.}
The \A\ performs every cautious step. The \JointLeader\ agents remain at
its step pebble. When the \A\ returns, it retrieves the pebble and the
entire group makes the final clockwise crossing to the node just
certified safe. The next cautious step then begins.

A group is \emph{available} when it is not committed to an unfinished
cautious step. Available groups that meet, merge and assign the roles above.
Groups committed to different unfinished steps first complete their own
returns.

 If an agent or group reaches a step pebble at \(u\)
whose owner is absent, it waits at \(u\) and does not remove the pebble.
The owner is either at the clockwise neighbour \(u^+\), trying to
return, or entered the \BH\ there. A returning owner retrieves the
pebble and joins the waiting agents. If \((u,u^+)\) is present in round
\(r\) but no owner reaches \(u\) at the beginning of round \(r+1\), a
waiting agent correctly identifies \(u^+\) as the \BH.

\paragraph{The Phase~1 boundary.}
Phase~1 ends after round \(6n\), and Phase~2 begins in round \(6n+1\). The Phase~1 boundary is the end of round $6n$.
The {\em boundary} configuration is the configuration at the beginning of
round \(6n+1\), after processing all traversals attempted in round
\(6n\) and retrieving the pebbles of completed returns, but before any
Phase~2 movement is chosen.
Consider an execution in which no agent has already terminated knowing
the footprint.
Lemma~\ref{lemma:variationgathering} proves that survivors from Phase 1 are in one
of the following configurations, for some node \(u\) and its clockwise neighbour
\(u^+\):
\begin{enumerate}
    \item all surviving agents are co-located at \(u\) (Figure~\ref{fig:phase1-configurations}.a); or
    \item one agent is at \(u^+\), attempting to return
    counter-clockwise, its pebble is on \(u\), and every other
    surviving agent is at \(u\) (Figure~\ref{fig:phase1-configurations}.b).
\end{enumerate}
In the second configuration, the probing agent is completing the return of the
group's last cautious step. Figure~\ref{fig:phase1-configurations} shows
the two possibilities.

\begin{figure}[htbp]
\centering
\begin{subfigure}[t]{0.485\textwidth}
\centering
\begin{tikzpicture}[x=.9cm,y=.82cm,line cap=round,line join=round]
    \path[use as bounding box] (-.65,-.55) rectangle (5.65,1.45);
    \foreach \x in {0,...,5}
        \node[glnode] (c\x) at (\x,0) {};
    \draw[gledge] (c0)--(c1)--(c2)--(c3)--(c4)--(c5);
    \draw[gledge,densely dotted] (-.55,0)--(c0);
    \draw[gledge,densely dotted] (c5)--(5.55,0);
    \node[below=2pt of c2,font=\scriptsize] {\(u\)};
    \node[glagent] (all) at ($(c2)+(0,.9)$) {};
    \node[right=2pt of all,font=\scriptsize]
        {2 or 3 agents};
    \draw[gledge] (all.south)--(c2.north);
\end{tikzpicture}
\caption{All survivors are at \(u\).}
\end{subfigure}
\hfill
\begin{subfigure}[t]{0.485\textwidth}
\centering
\begin{tikzpicture}[x=.9cm,y=.82cm,line cap=round,line join=round]
    \path[use as bounding box] (-.65,-.55) rectangle (5.65,1.45);
    \foreach \x in {0,...,5}
        \node[glnode] (s\x) at (\x,0) {};
    \draw[gledge] (s0)--(s1)--(s2);
    \draw[gledge,densely dashed] (s2)--(s3);
    \draw[gledge] (s3)--(s4)--(s5);
    \draw[gledge,densely dotted] (-.55,0)--(s0);
    \draw[gledge,densely dotted] (s5)--(5.55,0);
    \node[below=2pt of s2,font=\scriptsize] {\(u\)};
    \node[below=2pt of s3,font=\scriptsize] {\(u^+\)};
    \node[glagent] (others) at ($(s2)+(0,.9)$) {};
    \node[left=1pt of others,font=\scriptsize]
        {1 or 2 agents};
    \draw[gledge] (others.south)--(s2.north);
    \node[glagent] (ret) at ($(s3)+(0,.9)$) {\(a\)};
    \draw[gledge] (ret.south)--(s3.north);
    \draw[glmove] ($(s3)+(-.03,.36)$) to[bend right=18]
        ($(s2)+(.03,.36)$);
    \node[glpebble] at ($(s2)+(-.14,.16)$) {};
\end{tikzpicture}
\caption{Agent \(a\) is returning from \(u^+\) to \(u\).}
\end{subfigure}
\caption{The two Phase~1 boundary configurations. Squares denote agents,
the black dot denotes a step pebble, and the dashed edge may be absent.}
\label{fig:phase1-configurations}
\end{figure}

If an agent entered the \BH\ during Phase~1, its pebble remains on the
counter-clockwise neighbour \(h^{-}\) of the \BH. We call this permanent
pebble the \emph{black-hole marker}.

\subsubsection{Phase 2: resolving the boundary configuration}

Phase~2 consists exactly of rounds
\(6n+1,\ldots,6n+T\). It dispatches each Phase~1 boundary configuration
to either \RT\ or \BackwardCP, and  any pending return of a cautious step is granted \(T\)
rounds to complete. We first describe the two procedures and then
specify how Phase~2 handles each of the two boundary configuration.

\paragraph{Procedure \RT.}
\noindent\emph{Roles.}
The procedure uses one \Leader, one \A, and one \R.

\smallskip
\noindent\emph{Purpose.}
\RT\ is the known three-agent procedure used when all three agents are
co-located.

\smallskip
\noindent\emph{Operation.}
The \Leader\ and \A\ explore clockwise by probe--return--final-crossing
cycles, while the \R\ tests progressively farther nodes
counter-clockwise and reports after each test. These cycles place no
pebble: the waiting \Leader\ supplies the return evidence. The
failed-report test detects a destroyed \R. The complete procedure and
its guarantees are given in
Section~\ref{sec:cautious-pendulum} and Theorem~\ref{th:pendulum}.

\paragraph{Procedure \BackwardCP.}
\noindent\emph{Roles.}
Two co-located agents assign the smaller-identifier agent the role
\AggressiveLeader\ and the other agent the role \R. In one invocation
below, a returning third agent may later join them and assume the role
\A.

\smallskip
\noindent\emph{Purpose.}
The procedure \RT\ cannot start with only two co-located agents.
\BackwardCP\ lets such a pair continue the search without allowing the
\AggressiveLeader\ to enter the \BH. It also acts as a bridge to \RT\
when a third agent is still completing a cautious return.

\smallskip
\noindent\emph{Legal starts.}
\GL\ invokes \BackwardCP\ only in the following configurations:
\begin{enumerate}
    \item \emph{Marked start:} the black-hole marker is the only pebble
    lying on a node and is at the pair's node or clockwise ahead of it,
    at \(h^{-}\);
    \item \emph{Waiting start:} the pair is at a node \(u\) containing
    the step pebble of a surviving agent at \(u^+\) that is trying to
    return to \(u\).
\end{enumerate}

Figure~\ref{fig:backwardcp-invocations} illustrates the marked and waiting
starts and the corresponding movements.

\noindent\emph{Operation.}
The \AggressiveLeader\ maintains \(\ell\), its number of clockwise moves
since the start of the procedure. In a marked start it attempts to move
clockwise at every activation until it reaches the first pebble, which
is the black-hole marker; an absent edge merely blocks that attempt. In
a waiting start it is already at the first pebble and does not leave
\(u\). At a pebble it stays when the clockwise edge is present. \BackwardCP\ places no new pebble.

Concurrently, the \R\ performs the numbered departures of \RT: on
departure \(i\), it passes the \(i-1\) nodes tested earlier, tests the
next counter-clockwise node, and returns to report. The
\AggressiveLeader\ applies the same failed-report test as the \Leader\
of \RT.

At either legal start, suppose the \AggressiveLeader\ is at a pebble on
\(w\). If \((w,w^+)\) is present in round \(r\), it expects the pebble
owner to arrive at the beginning of round \(r+1\). If no owner arrives,
the \AggressiveLeader\ identifies \(w^+\) as the \BH.

In a waiting start, a successful return makes the returning agent
retrieve the step pebble and assume the role \A; the \AggressiveLeader\
assumes the role \Leader; and the \R\ retains its role and current
departure. Its departure number and failed-report counter are preserved,
so the execution continues as \RT.

\begin{figure}[htbp]
\centering
\begin{subfigure}[t]{0.485\textwidth}
\centering
\begin{tikzpicture}[x=.9cm,y=.82cm,line cap=round,line join=round]
    \path[use as bounding box] (-.65,-1.45) rectangle (5.65,1.45);
    \foreach \i in {0,...,4}
        \node[glnode] (p\i) at (\i,0) {};
    \node[circle,draw,line width=.55pt,fill=black,
          minimum size=4.2mm,inner sep=0pt] (p5) at (5,0) {};
    \draw[gledge] (p0)--(p1)--(p2)--(p3)--(p4)--(p5);
    \draw[gledge,densely dotted] (-.55,0)--(p0);
    \draw[gledge,densely dotted] (p5)--(5.55,0);
    \node[below=2pt of p4,font=\scriptsize] {\(h^{-}\)};
    \node[below=2pt of p5,font=\scriptsize] {\BH};
    \node[glpebble] at ($(p4)+(-.14,.16)$) {};
    \node[glagent] (al1) at ($(p1)+(0,.92)$) {\(\mathsf{AL}\)};
    \node[glagent] (r1) at ($(p1)+(0,-.96)$) {\(\mathsf{R}\)};
    \draw[glmove] (al1.east) to[bend left=10] ($(p4)+(0,.27)$);
    \draw[glbimove] ($(p0)+(-.5,-.96)$)--(r1.west);
\end{tikzpicture}
\caption{The black-hole marker at \(h^{-}\) precedes the \BH\ clockwise.}
\end{subfigure}
\hfill
\begin{subfigure}[t]{0.485\textwidth}
\centering
\begin{tikzpicture}[x=.9cm,y=.82cm,line cap=round,line join=round]
    \path[use as bounding box] (-.65,-1.45) rectangle (5.65,1.45);
    \foreach \i in {0,...,5}
        \node[glnode] (t\i) at (\i,0) {};
    \draw[gledge] (t0)--(t1)--(t2);
    \draw[gledge,densely dashed] (t2)--(t3);
    \draw[gledge] (t3)--(t4)--(t5);
    \draw[gledge,densely dotted] (-.55,0)--(t0);
    \draw[gledge,densely dotted] (t5)--(5.55,0);
    \node[below=2pt of t2,font=\scriptsize] {\(u\)};
    \node[below=2pt of t3,font=\scriptsize] {\(u^+\)};
    \node[glpebble] at ($(t2)+(-.14,.16)$) {};
    \node[glagent] (al2) at ($(t2)+(0,.92)$) {\(\mathsf{AL}\)};
    \node[glagent] (r2) at ($(t2)+(0,-.96)$) {\(\mathsf{R}\)};
    \node[glagent] (ret2) at ($(t3)+(0,.92)$) {\(a\)};
    \draw[glbimove] ($(t0)+(-.5,-.96)$)--(r2.west);
    \draw[glmove] ($(t3)+(-.03,.36)$) to[bend right=18]
        ($(t2)+(.03,.36)$);
\end{tikzpicture}
\caption{Agent \(a\) is returning to its pebble on \(u\).}
\end{subfigure}
\caption{The marked start (a) and waiting start (b) of \BackwardCP.
Here \(\mathsf{AL}\) and \(\mathsf{R}\) denote \AggressiveLeader\ and
\R, respectively. A small black dot denotes a pebble, the large filled
node is the \BH, and the dashed edge may be absent.}
\label{fig:backwardcp-invocations}
\end{figure}

In a marked start, the first pebble stops the \AggressiveLeader\ at
\(h^{-}\); in a waiting start, the step pebble keeps it at \(u\).
Consequently it cannot enter the \BH\ in either configuration.
Lemma~\ref{lemma:backwardcp} proves the correctness of both termination
tests and the bound of fewer than \(4n^2\) rounds.

\paragraph{Phase~2 dispatch.}
If no agent has already terminated, the two boundary configurations from Phase 1 are
handled as follows.

\begin{description}
    \item[Three agents co-located at \(u\).]
    They start \RT.

    \item[Exactly two surviving agents co-located at \(u\).]
    They start \BackwardCP\ in its marked configuration; the black-hole
    marker is at their node or lies clockwise ahead of them.

    \item[Three-survivor split: Two agents at \(u\), one returning from \(u^+\).]
    The pair at \(u\) begins the waiting start described above, while the
    third agent continues its return attempts. The pair uses no timeout.
    A successful return triggers the role transition described above and
    starts \RT; otherwise
    \BackwardCP\ terminates correctly within fewer than \(T\) rounds.

    \item[Two-survivor split; One agent at \(u\), one returning from \(u^+\).]
    Exactly two agents survive. The agent at \(u\) waits, and the other
    attempts to return in every Phase~2 round. A successful
    return makes the pair start \BackwardCP\ in its marked configuration.
    In this case the third agent entered the \BH\ during Phase~1, so its
    pebble is the black-hole marker required by the marked start.
    
\end{description}

The returning agent uses the same local rule whether zero, one, or two
agents are waiting at \(u\), because it cannot observe the remote
endpoint. A waiter uses the complementary local rule: if its clockwise
edge is present in round \(r\), it expects the owner to arrive at the
beginning of round \(r+1\). A completed return is processed before this
test; if no owner arrives, the waiter identifies the clockwise neighbour
as the \BH. In particular, a return completed in the last round of Phase~2 is
processed at the beginning of the next round before any timeout rule.
An otherwise isolated Phase~2 agent that is neither waiting at a pebble
nor completing a pending return stays in place through round \(6n+T\);
the Phase~1 boundary lemma (Lemma~\ref{lemma:variationgathering})
together with the dispatch cases implies that this can occur only after
another agent has already terminated correctly. 

\subsubsection{Phase 3: completing unresolved executions}

Phase~3 consists of rounds \(6n+T+1,\,6n+T+2,\ldots\) (here, we do not have a preset
last round). 
Its purpose is to finish any search that is still active at the Phase~2
timeout. If no correct termination has occurred earlier, let \(\tau\) be
the round of the first correct termination; we prove below that
\(\tau=\mathcal{O}(n^2)\). An execution of \RT\ or \BackwardCP\ already in
progress continues in Phase~3, which also uses the procedure \Forward.

\paragraph{Procedure \Forward.}
\noindent\emph{Role.}
The procedure uses one agent in the role \Scout.

\smallskip
\noindent\emph{Purpose.}
\Forward\ resolves the two-survivor split that remains when the
pending return has not completed by the Phase~2 timeout. It is also the
fallback for an otherwise isolated Phase~3 agent that is neither
executing another procedure nor completing a pending return, after some
agent has already terminated correctly.

\smallskip
\noindent\emph{Operation.}
A \Scout\ executing \Forward\ places no pebble and performs no cautious
step. At
each activation it first inspects its node. If the node contains a
pebble, it terminates knowing that the clockwise neighbour is the \BH.
If it meets another \Scout\ executing \Forward, the pair starts
\BackwardCP\ in a marked start.
Otherwise it attempts to move clockwise.

\paragraph{Phase~3 entry.}
First, the agents process a return completed in the last round of Phase~2;
every still-pending owner at \(u^+\) uses the same final rule, regardless
of how many agents are at the remote endpoint \(u\). If \((u,u^+)\) is
present in the first round of Phase~3, it makes one final return attempt and
retrieves its step pebble at the next activation. Afterwards, if
it meets two agents, it starts \RT;  if it meets
one, it starts \BackwardCP\ in a marked start; if it is alone, it starts \Forward\ as a \Scout.

If the edge is absent, a lone waiter at \(u\), if present, removes the
step pebble. Then, the waiter and the pending owner each start \Forward\ as
a \Scout. With no waiter, the owner starts \Forward\ as a \Scout\ at
\(u^+\), and its step pebble remains at \(u\).

By the Phase~1 boundary lemma and the Phase~2 dispatch, this
lone agent case can occur only after another agent has already correctly
terminated; hence, removing a pebble, that is in fact the
black-hole marker, does not compromise the required termination.

\subsubsection{Reference Implementation of the Algorithm}

A JavaScript reference implementation and simulator of the algorithm are
available at
\url{https://github.com/gadiluna/Gather-Locate/}. Furthermore, interested readers can also simulate the algorithm at:\\
\url{https://gadiluna.github.io/Gather-Locate/}.

\subsection{Correctness of \GL}\label{sec:correctness}

We first prove that at most one agent enters the \BH\ during Phase~1
and characterize the positions of the surviving agents after \(6n\)
rounds.

\begin{lemma}\label{scattered:goodmarking}
During Phase~1, at most one agent enters the \BH.  If an agent enters
the \BH, it enters clockwise from the counter-clockwise neighbour
\(h^{-}\) of the \BH\ and leaves its pebble on \(h^{-}\), where it
remains.
\end{lemma}

\begin{proof}
The only Phase~1 movement that can enter the \BH\ is the first
clockwise crossing \(u\to u^+\) of a cautious step. Immediately before
that crossing, the moving agent leaves its pebble on \(u\). The only
counter-clockwise Phase~1 movements are cautious-step returns
\(u^+\to u\). Their destination \(u\) was occupied immediately before
the first crossing and therefore is not the \BH.

Suppose that agent \(a\) enters the \BH.  It must therefore enter
clockwise from \(h^{-}\), after leaving its pebble there. During
Phase~1, only the owner of a step pebble retrieves it; another
agent or group that encounters the pebble is instead forced to wait. Since \(a\)
has been destroyed, its pebble is never removed.

Every surviving agent or group that later reaches \(h^{-}\) sees that
pebble and waits rather than beginning another clockwise probe.
Consequently no second agent can enter the \BH\ during Phase~1.
\end{proof}

\begin{observation}\label{scatter:correctterm}
Whenever an agent terminates at a step pebble because an observed present
edge is not followed by the owner's return, the agent knows the footprint. In
particular, every agent that terminates during Phase~1 knows the footprint.
\end{observation}

\begin{proof}
Let an agent wait at a node \(u\) containing a pebble whose owner is
absent. The owner previously probed clockwise from \(u\) to \(u^+\)
and, if it survived, attempts to return whenever \((u,u^+)\) is
present. Thus, if the edge is present in round \(r\), a surviving owner
reaches \(u\) at the beginning of round \(r+1\). Therefore, if no owner arrives at
that activation, it must have been destroyed at \(u^+\). The waiting
agent then knows that \(u^+\) is the \BH; because it knows \(n\), it
knows both oriented distances to the \BH\ and hence the footprint.
\end{proof}

The following lemma shows that the only possible configurations at round $6n$ are the ones shown in Figure~\ref{fig:phase1-configurations}.

\begin{lemma}[Phase~1 boundary configuration]
\label{lemma:variationgathering}
Let us assume that no agent terminates before the Phase~1 boundary. Then there
is a safe node \(u\) such that either every surviving agent is at \(u\),
or exactly one surviving agent is at \(u^+\), completing the
counter-clockwise return of a cautious step, while every other survivor
and the returning agent's step pebble are at \(u\).
\end{lemma}

\begin{proof}
Let \(H\) be the set of agents that survive through the Phase~1
boundary; by assumption, $|H| \leq 3$. Fix \(a\in H\), and let's focus on the group containing
\(a\) ({\em group} as defined by \JointCW). For each \(r\in\{1,\ldots,6n+1\}\), let
\(S_a(r)\subseteq H\) be the members
of that group after completed returns and meetings have been processed
at the beginning of round \(r\), but before movement is chosen. In general, an
agent {\em joins} a group when it either starts executing \JointCW\ with other two agents, or when it begins performing or
waiting for the same unfinished cautious step. Groups cannot
split, so \(S_a(r)\subseteq S_a(r+1)\).

For every round $r$ in Phase~1 where \(S_a(r)\neq H\), let \(e_a(r)\)
be the edge currently required by the group to progress: this edge is either the edge of its unfinished
cautious step, or, if one exists, its next clockwise edge.
Let
\[
 P_a=\bigl|\{r\in\{1,\ldots,6n\}:S_a(r)\neq H
       \text{ and }e_a(r)\in E_r\}\bigr|.
\]
Processing a return and assigning roles consume no additional round.
Hence every completed safe cautious step contributes at most three
rounds to \(P_a\), one for each traversal. The group advances along one
clockwise segment and can complete at most \(n-2\) safe steps before either
reaching the \BH, or another group, or a pebble. At most one additional
incomplete or fatal step contributes at most two rounds.
Therefore
\[
 P_a\leq 3(n-2)+2=3n-4.
\]

Suppose, by contradiction, that the group of \(a\) is still a
proper subset of \(H\) at the Phase~1 boundary, and choose
\(b\in H\setminus S_a(6n+1)\). Let's track the group of \(b\) similarly to what we have done for $a$. Because groups never split, the two tracked groups remain distinct
throughout Phase~1. Their required edges are therefore distinct in
every round: if both required \((u,u^+)\), their members would either be
co-located at \(u\), or one group would contain the unique agent
returning from \(u^+\) while the other waited at its pebble on \(u\).
In either case the grouping rules would merge them; thus, the required edges must be distinct.

Since \(P_a\leq3n-4\), the edge required by \(a\)'s group is absent in
at least \(6n-(3n-4)=3n+4\) rounds. In each such round every other ring
edge is present, including the distinct edge required by \(b\)'s group.
Thus \(P_b\geq3n+4\), contradicting the same upper bound
\(P_b\leq3n-4\).

Consequently, by the end of round \(6n\), all surviving agents perform
or wait for the same cautious step. If no return is pending, they are
co-located at its certified-safe node \(u\). Otherwise exactly one
agent, the pebble owner, is at \(u^+\) attempting the return, while its
step pebble and every other survivor are at \(u\). Because groups merge when
they meet at an endpoint of the same unfinished step, no separate group
can remain beside the one of the returning agent.
The node \(u\) is safe because the cautious step started there.
\end{proof}

\begin{observation}[Phase~1 pebble invariant]
\label{obs:phase1-pebbles}
Under the hypotheses of Lemma~\ref{lemma:variationgathering}, every
surviving agent carries its pebble except the returning agent in the
split configurations (see {\em Phase~2 dispatch}), whose step pebble is at \(u\). The only other
pebble that may lie on a node is the black-hole marker at \(h^{-}\).
\end{observation}

\begin{proof}
The owner retrieves the step pebble after every completed cautious step,
and no agent removes the pebble of an absent owner. By
Lemma~\ref{lemma:variationgathering}, at most one cautious step is
unfinished at the boundary. Lemma~\ref{scattered:goodmarking} accounts
for the only possible additional pebble, namely the marker left by the
single possible Phase~1 casualty.
\end{proof}

\begin{lemma}\label{lemma:phase1}
At the Phase~1 boundary, either an agent has terminated knowing the
footprint, or two or three surviving agents satisfy
Lemma~\ref{lemma:variationgathering} and
Observation~\ref{obs:phase1-pebbles}. If exactly two agents survive, the
third entered the \BH\ and left the black-hole marker at \(h^{-}\).
\end{lemma}

\begin{proof}
If an agent terminates, it knows the footprint by
Observation~\ref{scatter:correctterm}. Otherwise,
Lemma~\ref{scattered:goodmarking} leaves two or three survivors, and the
remaining claims follow from Lemma~\ref{lemma:variationgathering} and
Observation~\ref{obs:phase1-pebbles}.
\end{proof}

We next prove the bounds for Phases~2 and~3.

\begin{observation}[Compatibility of \BackwardCP\ and \RT]
\label{obs:bcp-to-rt}
When the returning agent reaches a pair executing \BackwardCP\ in a waiting start,
the role transition specified above produces a configuration reachable
in \RT.
\end{observation}

\begin{proof}
In this invocation of \BackwardCP, \AggressiveLeader\ remains at \(u\). We build a \RT\ execution 
with all three agents at \(u\), that in one round reaches the configuration of the first  \BackwardCP\ round.

If \((u,u^+)\) is absent, use one preliminary \RT\ round in which both
edges incident to \(u\) are present. Then \A\ reaches \(u^+\), \Leader\
remains at \(u\), and the hypothetical \R\ matches the first move of the
actual \R. Its counter-clockwise edge is necessarily present.
Neither \Leader\ nor \AggressiveLeader\ increments the failed-report
counter in this departure-initiation round. After this round the two
executions agree, apart from their role names and the inert step pebble,
so give the hypothetical execution the actual schedule beginning with
the second \BackwardCP\ round.

If \((u,u^+)\) is present in the first actual round, use a
preliminary \RT\ round in which that edge is present and the
counter-clockwise edge at \(u\) is absent. This places \A\ at \(u^+\)
while \Leader\ and \R\ remain at \(u\). The only temporary difference is the inert Phase~1 step pebble
at \(u\) in the actual execution; the returning agent removes it before
the transition.

Consequently, when \A\ returns, all positions, roles, departure data,
and counters match a configuration reachable in \RT\ immediately after
\A's successful return. Processing the return before any termination
test therefore allows the agents to continue that \RT\ execution.
\end{proof}

\begin{lemma}\label{lemma:backwardcp}
From any legal start, within fewer than \(4n^2\)
rounds, \BackwardCP\ either makes an agent terminate knowing the
footprint or, in a waiting start, receives the returning agent and
continues as \RT. Every agent that terminates knows the footprint.
\end{lemma}

\begin{proof}
We prove, in order, the safety of \AggressiveLeader, the validity of the
failed-report test, the round bound, and the correctness of every
termination.

In a marked start, until reaching \(h^{-}\), \AggressiveLeader\ remains
on the safe clockwise segment preceding \(h^{-}\); it attempts its next
edge and is blocked whenever that edge is absent, and stops at the first
pebble it encounters,
which by the hypothesis is the pebble on \(h^{-}\). In a waiting start,
it starts at \(u\) and does not cross
\((u,u^+)\). Thus \AggressiveLeader\ does not enter the \BH\ in either
case.

To bound the number of moves, cut the ring at the \BH; the \(n-1\) safe nodes
then form a path. Between a departure of \R\ and its next report, \R\
moves counter-clockwise to the next node not tested on an earlier
departure and then clockwise until it meets \AggressiveLeader. The
latter moves clockwise toward the pebble on \(h^{-}\) or remains at
\(u\). Neither agent crosses an end of the path of safe nodes. The
counter-clockwise part of a departure uses at most \(n-1\) edges, and
the clockwise return uses at most \(n-1\) edges. Consequently, from a
departure to its report, a surviving \R\ makes fewer than \(2n\) moves.
Each report follows the test of one additional safe node, so at most
\(n-2\) reports precede the departure on which \R\ tests the \BH.

We next verify the failed-report test. In each counted round \(r\), let
\(e_r\) be the clockwise edge incident to \AggressiveLeader's current
node. The counter is incremented precisely when
\AggressiveLeader's attempted traversal of \(e_r\) is blocked. Thus
\(e_r\) is absent and every other ring edge is present in that round.
Until it reports, a surviving \R\ lies at or strictly counter-clockwise
from \AggressiveLeader\ on the path of safe nodes. On its outward leg it
moves counter-clockwise. On its return leg, reaching
\AggressiveLeader's node produces a report before another traversal or
termination test is selected. Hence \R\ never attempts \(e_r\) before
reporting and makes one successful traversal in every counted round.

Fewer than \(2n\) such traversals suffice for a surviving \R\ to reach
the next untested node and return. A report produced by the final
traversal is processed at the next activation before the threshold is
tested. Therefore, if no report is present when the counter has reached
\(2n\), \R\ entered the \BH\ on its current departure.

Number departures from \(1\), and measure clockwise displacement from
the starting node of \BackwardCP\ as positive. On departure \(i\), \R\
tests the node of displacement \(-i\pmod n\). If
\AggressiveLeader's current displacement is \(\ell\), its clockwise
distance to the tested node is \((-i-\ell)\bmod n\), and its
counter-clockwise distance is \((\ell+i)\bmod n\). Thus \(i\),
\(\ell\), and \(n\) determine the footprint. Notice that this argument
does not require \(e_r\) to be incident to the \BH: a blocked
clockwise attempt can certify the progress of \R\ even when the node
tested by \R\ is several edges away.

We now bound the number of rounds by charging each round to an {\em event} (either a move or a round counted by the \AggressiveLeader).
Before the fatal departure, \R\ makes fewer than \(2n(n-2)\) moves.
The \AggressiveLeader\ makes fewer than \(n\) clockwise moves, and
fewer than \(n\) further activations process a report and initiate the
next departure. The fatal departure and its detection require at most
\((n-1)+2n+1=3n\) additional rounds.

Any earlier round not yet charged is one in which \R\ is blocked. In
such a round \AggressiveLeader's clockwise edge is present. If
\AggressiveLeader\ has not reached a pebble, it moves, and the round is
already charged to one of its fewer than \(n\) moves. If it is waiting
at a pebble, the following activation either processes the pebble
owner's return or identifies the clockwise neighbour as the \BH. This
terminal suffix contributes at most one activation beyond the preceding
allowances. Hence the total is less than
\[
  2n(n-2)+n+n+(3n+1)=2n^2+n+1<4n^2
\]
for \(n>3\).

Finally, if an agent reaches \(u\) from \(u^+\), it takes its pebble and
the agents handle that return before \AggressiveLeader\ checks whether
it can terminate.
Observation~\ref{obs:bcp-to-rt} shows that the agents then continue
according to \RT. If \((u,u^+)\) was present in round \(r\) and no
agent reaches \(u\) from \(u^+\) at the beginning of round \(r+1\),
the agent that left the pebble entered the \BH\ at \(u^+\). Hence,
whenever \AggressiveLeader\ terminates in \BackwardCP, it knows the
footprint.
\end{proof}

\begin{lemma}\label{lemma:two-agents-split}
From the two-survivor split configuration at the Phase~1 boundary, an
agent terminates knowing the footprint within \(\mathcal{O}(n^2)\)
rounds.
\end{lemma}

\begin{proof}
Use \(u\) and \(u^+\) as in Lemma~\ref{lemma:variationgathering}.
By Lemma~\ref{lemma:phase1}, the agent that entered the \BH\ left its
pebble at \(z=h^{-}\).
Since the agent at \(u^+\) survived, \(u^+\) is not the \BH, and therefore
\(u\neq h^{-}=z\). Thus the pebble on \(u\) is the returning agent's
step pebble and is distinct from the black-hole marker on \(z\).

Until a successful return occurs, the agent at \(u\) waits and the agent
at \(u^+\) attempts to return in every Phase~2 round. A successful return
makes the pair start \BackwardCP\ in a marked start, so
Lemma~\ref{lemma:backwardcp} applies. A return completed in the last
Phase~2 round is processed before Phase~3 entry. If no return has occurred
by then and the edge is present in the first Phase~3 round, the prescribed
final return attempt succeeds and starts \BackwardCP.

Assume instead that the agents remain separated and the edge is absent
at Phase~3 entry. The waiting agent removes the step pebble at \(u\),
and both agents execute \Forward\ in the role \Scout. The only pebble
still lying on a node is then the black-hole marker on \(z\).

Until an agent reaches \(z\), let \(d_i\in\{0,\ldots,n-1\}\) be the
clockwise distance from agent \(i\) to \(z\), and set
\(\Phi=d_1+d_2\). An agent inspects its node before moving, so it
terminates upon reaching \(z\) and never moves clockwise past it. While
the agents occupy distinct nodes and neither is at \(z\), they attempt
distinct clockwise edges. At most one of those edges is absent, so at
least one agent moves and \(\Phi\) decreases by the number of moves.
Since initially \(\Phi\leq2(n-1)\), within at most \(2(n-1)\) rounds an
agent reaches \(z\) or the agents become co-located. In the latter case
\(z\) still lies ahead of their common node, so the marked-start
hypothesis of
Lemma~\ref{lemma:backwardcp} applies. The total time is
\(\mathcal{O}(n^2)\).
\end{proof}

\begin{lemma}\label{lemma:three-agents-split}
From the three-survivor split configuration at the Phase~1 boundary, an
agent terminates knowing the footprint within \(\mathcal{O}(n^2)\)
rounds.
\end{lemma}

\begin{proof}
Use \(u\) and \(u^+\) as in Lemma~\ref{lemma:variationgathering}.
In the first Phase~2 round, the two agents at \(u\) start \BackwardCP\
in a waiting start, while the third continues trying to return. By
Lemma~\ref{lemma:backwardcp}, within
fewer than \(4n^2\) rounds either an agent terminates correctly or the
return occurs. In the latter case,
Observation~\ref{obs:bcp-to-rt} converts the execution to \RT, and
Theorem~\ref{th:pendulum} completes the proof.
\end{proof}

\begin{lemma}\label{lemma:post-phase1}
Starting from any outcome of Lemma~\ref{lemma:phase1}, either an agent
has already terminated knowing the footprint, or an agent does so during
Phases~2 and~3 within \(\mathcal{O}(n^2)\) additional rounds. Every agent
that terminates knows the footprint.
\end{lemma}

\begin{proof}
If three agents are co-located, they execute \RT, and the claim follows
from Theorem~\ref{th:pendulum}. If exactly two agents survive and are
co-located, they execute \BackwardCP. Phase~1 movement is clockwise,
and after the black-hole marker is placed no survivor crosses
\(h^{-}\). Hence \(h^{-}\) and its unique placed pebble lie on the
clockwise segment from \AggressiveLeader's current node to the \BH, so
Lemma~\ref{lemma:backwardcp} applies. If the two survivors occupy the
two nodes of an unfinished cautious step,
Lemma~\ref{lemma:two-agents-split} applies.

If all three agents are surviving in the two-node configuration,
Lemma~\ref{lemma:three-agents-split} applies. If an agent already
terminated in Phase~1, it knows the footprint by
Observation~\ref{scatter:correctterm}. The same observation covers any
later termination by a waiter using the present-edge/missing-return test.
Every agent that terminates in \BackwardCP\ knows the footprint by
Lemma~\ref{lemma:backwardcp}, and the same holds for \RT\ by
Theorem~\ref{th:pendulum}.

These arguments do not assume that a missing edge eventually
reappears. A permanently missing safe edge that blocks a \Leader\ or
\AggressiveLeader\ is covered by the failed-report test. On the other hand, if it
separates two \Scout\ agents executing \Forward, the potential argument in
Lemma~\ref{lemma:two-agents-split} still applies because at most one
of their two distinct clockwise edges is absent.

The preceding cases already guarantee that some agent terminates
correctly. It remains to verify the safety requirement that an
additional agent executing \Forward\ as a \Scout\ in Phase~3 cannot later
terminate with an incorrect footprint. Terminated agents take no further
part in meeting or role-assignment rules. Let \(x\) be the node at which
the additional agent starts \Forward. Any
pebble that this agent previously left and did not recover is at a node
\(p\) from which the agent moved clockwise on the first crossing of a
cautious step. Before starting \Forward, such an agent either waited at
its Phase~1 boundary node or remained the owner of a pending return; it
did not execute \RT\ or \BackwardCP. That pebble therefore lies behind its current position in the
clockwise traversal, and Phase~1 never carries a survivor clockwise
past the \BH. Starting from \(x\), the agent therefore reaches the \BH\
before it can return to \(p\).

No procedure used after Phase~1 places a new pebble. An isolated agent
waiting at a foreign pebble applies the present-edge/missing-return test.
If its clockwise edge is present, the return is processed first at the
next activation; if no owner arrives, the agent terminates correctly by
Observation~\ref{scatter:correctterm}. If the edge is absent at the
Phase~3 boundary, the agent removes the pebble before starting \Forward.
If that pebble is the black-hole marker, another agent has already
terminated correctly, the other original agent is the marker's destroyed
owner, and no second active agent remains. The isolated \Scout\ then sees
no foreign pebble at which it could terminate incorrectly; it can only
remain blocked or enter the \BH.

In every other branch, the Phase~3 entry rule ensures that any pending
step pebble at a waiting endpoint is recovered or removed before
starting \Forward. By the return-priority rule,
a surviving owner recovers its step pebble before any meeting or
termination rule; if its return remains blocked until Phase~3, that
owner is the \Scout\ itself and its pebble is the pebble \(p\)
already considered. Lemma~\ref{scattered:goodmarking} accounts for the
only destroyed owner and its marker on \(h^{-}\). Hence a different pebble that the agent
encounters before the \BH\ can only be the marker on \(h^{-}\). Thus, if the
\Scout\ terminates while executing \Forward\ upon seeing a pebble, it is at
\(h^{-}\) and knows the footprint. It may instead remain blocked or
enter the \BH, neither of which affects the agent that has already
terminated knowing the footprint.
\end{proof}

\begin{theorem}\label{th:scat}
Given a dynamic oriented ring \({\cal R}\) of size \(n>3\), three
scattered agents with distinct visible identifiers and pebbles running \GL\
solve \BHS\ using \(\mathcal{O}(n^2)\) moves and
\(\mathcal{O}(n^2)\) rounds.
\end{theorem}

\begin{proof}
Phase~1 lasts \(6n=\mathcal{O}(n)\) rounds.  By
Lemma~\ref{lemma:phase1}, it either makes an agent terminate knowing the
footprint or produces one of the configurations handled by Phase~2.
Phase~2 lasts exactly \(T=4n^2=\mathcal{O}(n^2)\) rounds. If Phase~3 is
needed, it starts in round \(6n+T+1\); the correctness and the combined
\(\mathcal{O}(n^2)\)-round bound for Phases~2 and~3 follow from
Lemma~\ref{lemma:post-phase1}.

Every agent that terminates knows the footprint, and at least one agent
survives and terminates with that knowledge.
Since there are only three agents and each performs at most one move
per round, the cost through the first such termination is also
\(\mathcal{O}(n^2)\).
\end{proof}

Combining Theorem~\ref{madebyus0}, Theorem~\ref{scat:lb}, and
Theorem~\ref{th:scat} yields the following result.

\begin{theorem}\label{th:scatfinal}
Algorithm \GL\ is size-optimal, and its
\(\Theta(n^2)\) cost and time are asymptotically optimal among
size-optimal algorithms.
\end{theorem}

\section{Impossibility with Endogenous Communication}
\label{sec:scattered}

We conclude with an impossibility result for endogenous
communication. In the preceding sections agents communicate through
exogenous mechanisms supplied by the system, namely pebbles or
whiteboards. Endogenous mechanisms instead use only the agents'
capabilities. In the \emph{FaceToFace (F2F)} model agents communicate
only when they occupy the same node; in the \emph{Vision} model they
can see, but not communicate with, agents at their current node.

Three scattered agents equipped with pebbles suffice to solve \BHS.
Without an external communication object, even the stronger F2F model
does not suffice for three scattered agents, already on static rings.

\begin{theorem}\label{th:three-f2f-static}
No deterministic algorithm solves \BHS\ with three initially scattered
agents in the F2F model on every anonymous oriented static ring. More
precisely, for every deterministic algorithm there is a scattered
placement and a \BH\ position on an oriented static ring of size \(30\)
for which the algorithm fails. This holds even when the agents have
distinct visible identifiers and know that the ring size is \(30\).
\end{theorem}

\begin{proof}
Assume, for a contradiction, that a deterministic F2F algorithm \(A\)
solves \BHS\ for every scattered placement and every \BH\ position on
an oriented static ring of size \(30\) (the argument can be easily extended to any larger ring). Write its nodes as ${\cal R}=(v_0,v_1,\ldots,v_{29})$,
with indices modulo \(30\). These
indices are used only in the proof and are not visible to the agents.
As long as an agent neither meets another agent nor enters the \BH,
every snapshot contains both incident edges and no other agent. Since
\(A\) is deterministic, its identifier, the known size \(30\), and the
content of its local memory determine its next action.

For each of the three agents, consider the sequence of actions prescribed by \(A\) 
when every snapshot contains both incident edges and no other agent (i.e., the agent is alone),
and every node visited by the agent is safe. Let \(r_1\) be the first
round in these  sequences in which an agent either moves or terminates.
First, note that such a round must exist. Otherwise, place the agents on three distinct safe
nodes; since they never move and no agent terminates, \BHS is never solved.

Suppose that at least one agent terminates in round \(r_1\). This case
also includes the possibility that another agent moves in the same
round. By definition, before round \(r_1\), no agent has moved. Place the agents on
three distinct nodes and let \(x\) be an agent that terminates. At each
of its activations, \(x\) has seen both incident edges and no other
agent. Consequently, the content of its local memory at termination is
the same for every position of the \BH\ outside the three initial
nodes. There are \(30-3=27\) possible such positions, and they give
different clockwise and counter-clockwise distances from \(x\)'s
current node to the \BH. The same local memory cannot determine the
correct distances for all \(27\) positions. Choose one for which it
does not; then \(x\) terminates without knowing the footprint of
\({\cal R}\), thus \BHS is not solved.

We may therefore assume that no agent terminates in round \(r_1\) and
that at least one agent moves. Let \(a\) be an agent that moves in that
round, and let \(\sigma_a=+1\) if it moves clockwise and
\(\sigma_a=-1\) if it moves counter-clockwise. For the remainder of the
proof, without loss of generality, place the \BH\ at \(v_0\) and place \(a\) initially at
\(v_{-\sigma_a}\). Its move in round \(r_1\) brings $a$ to \(v_0\), so
\(a\) is destroyed by the \BH.

Let \(b\) and \(c\) be the other two agents. For \(x\in\{b,c\}\), define
\(e_x=+1\) if \(x\) moves clockwise in round \(r_1\), \(e_x=-1\) if
it moves counter-clockwise, and \(e_x=0\) if it stays still. Thus, if
\(x\) is at \(v_{q_x}\) at the beginning of round \(r_1+1\), its
initial node is \(v_{q_x-e_x}\). The action of \(x\) in round \(r_1\)
and the content of its local memory at the beginning of round
\(r_1+1\) do not depend on \(q_x\), because nodes are anonymous and
\(x\) has seen both incident edges and no other agent.

For each \(x\in\{b,c\}\), start with its local memory at the beginning
of round \(r_1+1\) and consider the following isolated continuation:
at every activation its snapshot contains both incident edges and no
other agent, and every edge that it attempts is traversed. Set
\(p_x(0)=0\). Until \(x\) terminates, let \(p_x(t)\) be the number of
clockwise moves minus the number of counter-clockwise moves made in the
first \(t\) rounds of this continuation (i.e., it represent the absolute distance they moved since round $r_1+1$). If \(x\) starts this
continuation at \(v_{q_x}\), then after \(t\) rounds it is at
\(v_{q_x+p_x(t)}\).

Let \(\tau\) be the smallest \(t\geq 0\) such that either one of the
two agents terminates at the beginning of round \(r_1+1+t\), or
\(\lvert p_x(t)\rvert=3\) for some \(x\in\{b,c\}\). If both conditions
hold for the same \(t\), we consider the case in which an agent
terminates.

If no such \(\tau\) exists, then
\(p_b(t),p_c(t)\in\{-2,-1,0,1,2\}\) for every \(t\), and neither
agent terminates. Place \(b\) at \(v_8\) and \(c\) at \(v_{-8}\) at
the beginning of round \(r_1+1\); their initial nodes are respectively
\(v_{8-e_b}\) and \(v_{-8-e_c}\). The first belongs to
\(\{v_7,v_8,v_9\}\), while the second belongs to
\(\{v_{-9},v_{-8},v_{-7}\}\). These sets are disjoint modulo \(30\)
and contain neither \(v_0\) nor \(v_{-\sigma_a}\). Moreover, the two
sets of nodes $v_6,v_7,\ldots,v_{10}$ and $v_{-10},v_{-9},\ldots,v_{-6}$ are respectively the residues \(6,\ldots,10\) and \(20,\ldots,24\);
they are disjoint and contain neither \(v_0\) nor
\(v_{-\sigma_a}\).
Therefore \(b\) and \(c\) never meet, never enter the \BH, and never
terminate. Since \(a\) has already been destroyed, \(A\) does not
solve \BHS.

Suppose now that an agent \(x\in\{b,c\}\) terminates at the beginning
of round \(r_1+1+\tau\), and let \(y\) be the other agent. For every
\(t\leq\tau\), both \(p_x(t)\) and \(p_y(t)\) belong to
\(\{-3,-2,\ldots,3\}\). Place \(y\) at \(v_{-10}\) at the beginning
of round \(r_1+1\), and consider two placements for \(x\): at \(v_8\)
and at \(v_9\). In the first placement, \(x\) visits only nodes among
\(v_5,\ldots,v_{11}\) before terminating; in the second, it visits
only nodes among \(v_6,\ldots,v_{12}\). Agent \(y\) visits only nodes
among \(v_{-13},\ldots,v_{-7}\), whose residues modulo \(30\) are
\(17,\ldots,23\). The nodes visited by \(x\) have residues between
\(5\) and \(12\), so the two visited sets are disjoint and contain
neither \(v_0\) nor \(v_{-\sigma_a}\). The possible initial nodes of
\(x\) have residues between \(7\) and \(10\), whereas the initial node
of \(y\) has residue \(19\), \(20\), or \(21\). Thus the initial nodes
are distinct and safe in both placements. Agent \(x\) has the same
snapshots and the same content of its local memory in the two
placements. Nevertheless, the correct clockwise and counter-clockwise
distances from \(x\)'s current node to \(v_0\) differ between them.
Hence \(x\) cannot know the
footprint of \({\cal R}\) in both placements, hence \BHS is not solved.

The only remaining case is that no agent terminates at the beginning
of round \(r_1+1+\tau\), and an agent, say \(b\), satisfies
\(p_b(\tau)=3\sigma\), where \(\sigma\in\{-1,+1\}\). Place \(b\) at
\(v_{-3\sigma}\) at the beginning of round \(r_1+1\). For
\(t<\tau\), \(\lvert p_b(t)\rvert<3\), so \(b\) does not enter
\(v_0\) before completing the first \(\tau\) rounds after round
\(r_1\). The move that produces
\(p_b(\tau)=3\sigma\) takes \(b\) to \(v_0\), where it is destroyed.
Its initial node is \(v_{-3\sigma-e_b}\). If \(\sigma=+1\), this node belongs to
\(\{v_{26},v_{27},v_{28}\}\); if \(\sigma=-1\), it belongs to
\(\{v_2,v_3,v_4\}\). In either case it is distinct from \(v_0\) and
from the initial node \(v_{-\sigma_a}\in\{v_1,v_{29}\}\) of \(a\).
All indices
used for the positions of \(b\) up to this point lie between \(-6\)
and \(6\). The difference between two such indices has absolute value
at most \(12<30\), so two of them represent the same ring node only
when they are equal.

It remains to choose the node occupied by \(c\) at the beginning of
round \(r_1+1\). Let this node be \(v_q\). For every
\(t\leq\tau\), both \(p_b(t)\) and \(p_c(t)\) belong to
\(\{-3,-2,\ldots,3\}\). Work in the residue set
\(\mathbb{Z}_{30}\) and define
\[
\begin{aligned}
B&=\{-p_c(t)\bmod 30:0\leq t\leq\tau\},\\
M&=\{-3\sigma+p_b(t)-p_c(t)\bmod 30:0\leq t\leq\tau\},\\
I&=\{e_c,\ e_c-\sigma_a,\ e_c-3\sigma-e_b\}.
\end{aligned}
\]
If \(q\in B\), agent \(c\) visits \(v_0\) by time \(\tau\). If
\(q\in M\), it meets \(b\) by that time. If \(q\in I\), its initial
node \(v_{q-e_c}\) equals \(v_0\), the initial node
\(v_{-\sigma_a}\) of \(a\), or the initial node
\(v_{-3\sigma-e_b}\) of \(b\), respectively.

Since \(p_c(t)\) takes values in a seven-element set,
\(|B|\leq 7\). Since \(p_b(t)-p_c(t)\in\{-6,-5,\ldots,6\}\),
\(|M|\leq 13\), and clearly \(|I|\leq 3\). Therefore the set
$Q=\mathbb{Z}_{30}\setminus(B\cup M\cup I)$ satisfies \(|Q|\geq30-(7+13+3)=7\). Thus, if $c$ is placed on one of the nodes \(q\in Q\), the three
agents start on distinct safe nodes, agents \(a\) and \(b\) are
destroyed by \(v_0\), and agent \(c\) has met no other agent.

Consider the moves of $c$ after the first \(\tau\)
rounds following round \(r_1\), while every node it visits is safe. If
\(c\) never terminates, then for each
\(q\in Q\) it either eventually enters \(v_0\), leaving
no surviving agent, or survives forever without terminating. In either case \(A\)
does not solve \BHS. 

Suppose instead that \(c\) eventually terminates.
If, for some \(q\in Q\), it enters \(v_0\) before
terminating, choose that value: all three agents are destroyed. If this
never happens, \(c\) follows the same isolated continuation and
terminates after the same number of rounds for every \(q\in Q\), with
the same content of its local memory. Let \(d\in\mathbb{Z}_{30}\) be
its total clockwise displacement at that activation. Its node is then
\(v_{q+d}\). The map \(q\mapsto q+d\bmod 30\) is injective, so the
correct clockwise and counter-clockwise distances from \(c\)'s current
node to \(v_0\) are different for distinct values of \(q\). The same
local memory can therefore represent the footprint correctly for at
most one \(q\in Q\). Since \(|Q|\geq7\), choose another value; for that
placement, \(c\) terminates without knowing the footprint of
\({\cal R}\).

In every case there is a static ring and a scattered placement on
which \(A\) does not solve \BHS, a contradiction.
\end{proof}

This impossibility contrasts with the co-located case, where three
agents solve \BHS\ in the same communication model
\cite{DiFPS25}. It remains open whether more than three scattered
agents suffice. We conjecture that
Theorem~\ref{th:three-f2f-static} extends to every constant number of
agents.

\section{Concluding Remarks}

We studied Black Hole Search in oriented $1$-interval connected dynamic rings
when the agents are \emph{scattered}, i.e., placed at distinct arbitrary safe
nodes rather than gathered at a common safe node. Working in the exogenous
communication setting, we gave a tight characterization of size-optimal
solutions. Two agents do not suffice, so three agents are size-optimal; for
three agents we proved an $\Omega(n^2)$ lower bound on both moves and rounds
(Theorem~\ref{scat:lb}) that holds even in the stronger whiteboard model and
even when agents carry distinct visible identifiers. We then presented
\textsc{Gather\&Locate}, a three-phase algorithm in the weaker pebble model
that matches this bound with $O(n^2)$ moves and $O(n^2)$ rounds (Theorem~\ref{th:scat}), and is therefore size-optimal with asymptotically
optimal cost and time (Theorem~\ref{th:scatfinal}). Finally, we showed that an
exogenous marking mechanism is not a convenience but a necessity: in the
endogenous \textsc{FaceToFace} model, three scattered agents cannot solve
\textsc{Bhs} even on a \emph{static} oriented ring of size $30$
(Theorem~\ref{th:three-f2f-static}).

Taken together, these results isolate the individual and combined costs of the
two features that distinguish our setting from earlier work. On a static ring,
two agents locate the black hole with $O(n)$ moves and rounds. Dynamicity
alone raises the requirement to three agents and, for co-located agents with
pebbles, to $O(n^{1.5})$ moves and rounds~\cite{DiFPS25}. Scattering the agents, on top
of dynamicity, raises the complexity further to $\Theta(n^2)$. The quadratic
lower bound is, in this sense, the price of not knowing where the other agents
are.

Several questions remain open. The most immediate is \emph{unoriented} dynamic
rings with scattered \emph{anonymous} agents. Both features we relied on,
orientation and distinct visible identifiers, are used pervasively in
\textsc{Gather\&Locate}, and removing them appears to demand genuinely
different techniques; even the solvability threshold is unclear. A second
direction is the effect of \emph{team size}. Our optimality is with respect to
the minimum number of agents, but larger teams could plausibly reduce the
$\Theta(n^2)$ cost, and identifying the optimal move/round trade-off as a
function of the number of agents is open.

A further direction might be also related to the use of \emph{randomization}. All algorithms in this
paper, and the lower bounds against which they are matched, concern
deterministic agents. Randomization could help in two distinct ways. First,
against an \emph{oblivious} adversary that fixes the edge-removal schedule in
advance, randomized cautious exploration might let scattered agents gather
faster, or break the symmetry that the deterministic $\Omega(n^2)$ argument
exploits, potentially lowering the expected move and round complexity below the
deterministic bound.

\section*{Acknowledgments} A preliminary version of this paper has been presented at the 27th Conference on Principles of Distributed Systems (OPODIS 2023). This research has been supported in part by NSERC under the Discovery Grant program, and by the Italian National Group for Scientific Computation GNCS-INdAM.

\bibliography{bh_scattered_dmtcs_2024}

\end{document}